\documentclass[trackchanges,twocolumn]{aastex701}
\usepackage{todonotes}

\begin{document}

\title{The HAges Catalog: Stellar Ages for High Priority HWO Target Stars}

\author[0000-0003-2828-0334]{Austin Ware}
\affiliation{School of Earth and Space Exploration\\Arizona State University\\Tempe, AZ 85287, USA}
\email{austin.t.ware@gmail.com}

\author[0009-0008-3554-7205]{Katelyn Ruppert}
\affiliation{School of Earth and Space Exploration\\Arizona State University\\Tempe, AZ 85287, USA}
\email{}

\author[0000-0003-1705-5991]{Patrick Young}
\affiliation{School of Earth and Space Exploration\\Arizona State University\\Tempe, AZ 85287, USA}
\email{}

%% Use the \collaboration command to identify collaborations. This command
%% takes an optional argument that is either a number or the word "all"
%% which tells the compiler how many of the authors above the command to
%% show. For example "\collaboration[all]{(DELVE Collaboration)}" wil include
%% all the authors above this command.
%%
%% Mark off the abstract in the ``abstract'' environment. 
\begin{abstract}

Precise stellar ages (uncertainties $\lesssim 1$ Gyr, or $\sim 20\%$ at solar age) are required to discern evolutionary trends in atmospheric biosignatures of terrestrial habitable zone exoplanets surveyed by the Habitable Worlds Observatory (HWO) and will aid in constraining planetary interior evolution and target prioritization. We present a catalog of stellar ages for Tier 1 and Tier 2 targets in the HWO Target Stars and Systems (TSS) sub-working group’s TSS25 list, compiling published literature ages derived from high-precision methods. The sample comprises 659 stars likely to be observed by HWO, independent of the final mission architecture. This initial catalog focuses on asteroseismology and gyrochronology, which can achieve $\sim 20\%$ precision for the majority of these stars. We find that only $\sim 5\%$ of the sample have asteroseismic ages and $\sim 20\%$ have gyrochronal ages, with just $\sim 2\%$ having constraints from both methods. For stars with multiple published measurements, the median reported statistical uncertainties are slightly smaller than the systematic uncertainties: $\sim 9\%$ versus $\sim 12\%$ for asteroseismology and $\sim 16\%$ versus $\sim 18\%$ for gyrochronology. The scarcity of precise stellar ages in this sample highlights the need for a concerted effort to obtain robust age constraints in advance of HWO; this catalog is intended as a living resource that will be regularly updated in the lead-up to the mission.

\end{abstract}

%% Keywords should appear after the \end{abstract} command. 
%% The AAS Journals now uses Unified Astronomy Thesaurus (UAT) concepts:
%% https://astrothesaurus.org
%% You will be asked to selected these concepts during the submission process
%% but this old "keyword" functionality is maintained in case authors want
%% to include these concepts in their preprints.
%%
%% You can use the \uat command to link your UAT concepts back its source.
%\keywords{\uat{Asteroseismology}{} --- \uat{Gyrochronology}{} --- \uat{Stellar Ages}{}}

%% From the front matter, we move on to the body of the paper.
%% Sections are demarcated by \section and \subsection, respectively.
%% Observe the use of the LaTeX \label
%% command after the \subsection to give a symbolic KEY to the
%% subsection for cross-referencing in a \ref command.
%% You can use LaTeX's \ref and \label commands to keep track of
%% cross-references to sections, equations, tables, and figures.
%% That way, if you change the order of any elements, LaTeX will
%% automatically renumber them.

\section{Introduction}\label{sec:intro}

Measurements of stellar ages for potential Habitable Worlds Observatory (HWO) targets are crucial both prior to and after HWO launches. Ages will aid in prioritizing systems by the likelihood of hosting planets that can still maintain habitable conditions and host detectable life \citep{lisse2020,ware2025}. Precise stellar ages, with uncertainties $\lesssim 1$ Gyr ($\sim 20$\% at solar age), are needed to constrain the potential mantle degassing lifetimes of terrestrial planets \citep{unterborn2022} and for discerning evolutionary trends in detected biosignatures \citep{bixel2020}.

A key product of the HWO Target Stars and Systems (TSS) sub-working group was the TSS25 list, which categorized potential HWO targets into priority tiers based on their likelihood to be observed and the need for stellar characterization prior to the mission \citep{tuchow2025}. Tiers 1 and 2 contain the current 659 targets most likely to be observed by HWO, regardless of the final observatory architecture. A consistently updated catalog of ages from high precision methods for Tiers 1 and 2 will serve as useful resource for the community.

Measurement of stellar ages is notoriously difficult and certain age-dating methods are best depending on the stellar spectral type and evolutionary phase \citep{soderblom2010}. We focus here on two state-of-the-art methods: asteroseismology and gyrochronology. Asteroseismology, often considered the gold standard for age-dating field stars, utilizes oscillations presenting as variations in brightness or radial velocity to constrain the interior structure and evolution \citep{kjeldsen1995}. Gyrochronology relates a star's measured rotation period to its age through a process known as magnetic braking. Cool stars with convective envelopes generate magnetic fields that couple to their ionized stellar winds. Magnetized winds carry away angular momentum along magnetic field lines, causing a star’s rotation to steadily slow over billions of years in a predictable manner \citep{skumanich1972}.

Obtaining precise ages through either asteroseismology or gyrochronology is possible for the predominantly FGK-type dwarfs in the top tiers of the TSS25 list. For asteroseismology, statistical precisions of $< 20$\% in age are common for well-characterized solar-like oscillators, with uncertainties of $\lesssim 10$\% possible for the best characterized stars \citep[e.g.,][]{valle2015,silvaaguirre2017,huber2022,li2024,pinsonneault2025}. Typical statistical uncertainties for ages derived via gyrochronology range between $\sim 10-40$\% \citep[e.g.,][]{barnes2007,mamajek2008,bouma2023,lu2024}, largely due to the empirical scatter of cluster rotation sequences combined with the stellar spin-down rate \citep{bouma2023}. Because the two techniques are based on fundamentally different approaches, they are subject to distinct sources of systematic uncertainties. Asteroseismology is model-dependent and relies on combinations of stellar evolution models, stellar oscillation models, and seismic scaling relations to infer stellar properties from observations. While solar calibration provides an important anchor, uncertainties in input physics lead to systematic errors that are often comparable to or larger than the quoted statistical precision \citep{valle2015,li2024}. For the Kepler dwarfs LEGACY sample \citep{silvaaguirre2017}, which is comprised of the best asteroseismically characterized main sequence stars to date, the median statistical and systematic age uncertainties are $\sim 8$\% and $\sim 12$\%, respectively.

In contrast, gyrochronology relies on empirical relations \citep[e.g.,][]{barnes2007,mamajek2008,angus2015,curtis2020,engle2023} or more physically motivated semi-empirical models \citep[e.g.,][]{gallet2013,matt2015,vansaders2016} calibrated to stellar clusters and field stars with well-determined ages. These calibrations inherit uncertainties from cluster age determinations, which are themselves model-dependent through main sequence turnoff fitting and often rely on stars more massive than the Sun \citep{soderblom2010,angus2015}. Gyrochronology also assumes that stellar rotation converges to a well-defined sequence with age, such that rotation provides a unique age diagnostic. While this convergence is observed at intermediate ages, young solar-type stars exhibit a significant spread in rotation rates that introduce intrinsic scatter and limit age precision \citep[e.g.,][]{angus2015,douglas2016,bouma2023}. In general, systematic uncertainties in gyrochronal ages are comparable to the statistical uncertainties, with dominant contributions from possible variations in the stellar spin-down rate with time and the absolute age scale calibration \citep{bouma2023}. Additional uncertainties arise from possible dependencies of the rotation–age relation on parameters such as metallicity, which remain poorly constrained \citep[e.g.,][]{gaidos2023}.

While both methods are applicable to Sun-like (FGK-type) main sequence stars, asteroseismology can also be applied to evolved solar-like oscillators \citep{jackiewicz2021} and classical pulsators \citep{kurtz2022}. Gyrochronology is typically limited to ages $\lesssim 4$ Gyr due to a lack of older calibration clusters \citep{bouma2023}, and its extension to older ages is further complicated by the observed flattening of the rotation–age relation first noticed in Kepler stars \citep{vansaders2016}. While younger, active stars follow the standard spin-down law, older stars exhibit anomalously fast rotation. This has been linked to a reduction in the large-scale magnetic field strength as the global stellar dynamo becomes less efficient \citep{metcalfe2017,metcalfe2025a}. However, calibrating gyrochronology relations with kinematic ages shows promise in extending gyrochronology to older ages \citep{lu2024}. Given that the bulk of the TSS25 Tier 1 and 2 stars are FGK-type dwarfs and subgiants, the combination of asteroseismology and gyrochronology will cover the vast majority of likely HWO targets.

We present here the HAges Catalog (pronunciation: hay-juhz or ha-guhs, like Scottland's national dish haggis), an initial catalog of published stellar ages for HWO Tier 1 and 2 stars from two high precision methods that will be applicable to most HWO targets: asteroseismology and gyrochronology. This catalog will highlight stars most in need of age measurements, provide summary statistics for stars with published ages, and provide detailed information on the individual measurements for each star.

We detail the construction of the HAges Catalog in Section \ref{sec:meth}. In Section \ref{sec:over}, we provide an overview of the current contents of the catalog. In Section \ref{sec:disc}, we discuss the typical systematic uncertainties for both age-dating methods and future prospects for obtaining precise ages.

\section{Compilation of the HAges Catalog}\label{sec:meth}

In this section, we describe the construction of the HAges Catalog, including a description of the adopted HWO target sample and the manner in which published ages were selected for each stellar age-dating method. 

\subsection{Sample}\label{sec:meth_samp}

The HWO TSS sub-working group published the community-developed TSS25 list of potential stellar targets for HWO to guide precursor and preparatory observations \citep{tuchow2025}. We adopt a subset of these targets for the HAges Catalog, which are summarized in Table \ref{tab:tss25} and detailed here. The TSS group started with the catalog of $\sim 13,000$ nearby ($< 50$ pc), bright ($T$ and $G < 12$) stars from the HWO Preliminary Input Catalog \citep[HPIC,][]{tuchow2024} as a basis. The distance and magnitude cutoffs were chosen to provide a volume complete sample of targets that could still be potentially suitable for a direct imaging exo-Earth search under the most optimistic of scenarios. The HPIC was then separated into three tiers according to both the current likelihood to be observed and the need for further observations to constrain stellar properties. Tier 1 contains the targets listed in the Exoplanet Exploration Program (ExEP) Mission Star List (EMSL). Taking into account the HZ limits for Earth \citep[0.95-1.67 AU,][]{kasting1993,Kopparapu2013}, exoplanet brightness, planet-star contrast ratio, the presence of disks, binarity, and hypothetical inner working angles for HWO, the EMSL reports 164 targets with HZs most accessible to a future direct imaging survey with a 6-m-class telescope. The goal of Tier 2 was to identify an additional set of targets with a high probability of being observed, regardless of the final mission architecture. Through exo-Earth yield simulations with the Altruistic Yield Optimization \citep[AYO,][]{ayo} and Exoplanet Open-Source Imaging Mission Simulator \citep[EXOSIMS,][]{exosims} tools, the TSS group selected 495 additional targets that make up Tier 2. All other targets in the HPIC were relegated to Tier 3. We adopt the combined 659 targets in Tiers 1 and 2 for the HAges Catalog, representing the current list of the most likely targets to be observed by HWO. This also keeps the sample at a manageable size that is suitable for the task of manually searching the literature for published ages. We expect that the likely sample of targets for HWO will change in the future as efforts are made to further characterize high priority targets and a final mission architecture is chosen. Should this happen, we will not remove any targets from the HAges Catalog, but will update HAges with any new additions. %A planet with a semimajor axis of 1 AU corresponds to a separation of 20 mas at 50 pc. Assuming an idealized coronograph with an inner working angle of $1 \lambda/D$, an aperture of 8m, and an operating wavelength of 1 $\mu m$ corresponding to the location of a prominent water feature, HWO would be limited to planets at separations $> 26$ mas. Therefore, outside 50 pc the habitable zone will typically fall inside the inner working angle. At 50 pc, the TESS Input Catalog becomes incomplete at $T \gtrsim 12$. Although targets with $T \gtrsim 8$ will likely be too faint for detecting exo-Earths, these targets may still be amenable to surveys of larger planets.

\begin{deluxetable}{ccl}
\tablewidth{\columnwidth}
\tablecaption{HAges Catalog Sample \label{tab:tss25}}
\tablehead{
\colhead{Sample} & \colhead{Stars} & \colhead{Description}
}
\startdata\\
TSS25 Tier 1 & 164 & \parbox[c]{0.55\columnwidth}{\raggedright Targets most accessible for an exo-Earth direct imaging mission} \\\\
TSS25 Tier 2 & 495 & \parbox[c]{0.55\columnwidth}{\raggedright Additional targets that could plausibly be observed by the range of proposed HWO designs} \\\\
\tableline
\textbf{Total} & 659 & \\
\enddata
%\tablecomments{}
\end{deluxetable}

\subsection{Asteroseismology}\label{sec:meth_seis}

Stellar ages derived through asteroseismology are often reported for only one or a limited number of stars in a publication. This necessitates a largely manual literature search rather than automated cross-matching of catalogs. We first queried the NASA Astrophysics Data System (ADS) using the unique SIMBAD identifiers for each Tier 1 and 2 target along with search terms related to asteroseismology. Since it can often take several months after publication for SIMBAD objects to be matched with a literature source, we did not require that the SIMBAD identifiers be present in papers after January 2025. We then manually searched the hundreds of publications returned by this query to identify stellar ages derived through asteroseismology. For each target present, we record the reported age along with the uncertainties and derived mass when available. When an age is reported for a star in a binary or multiple system, we only record the age for that component unless joint modeling is performed using properties from multiple components. We also record the implemented stellar evolution and oscillation codes to provide some context for the source of differences in reported ages and to give a brief overview of the common tools used.

In some cases, multiple pipelines are used within the same paper to derive the stellar age as a means to estimate the systematic uncertainty. Rather than only recording the adopted age, we record the results from all pipelines. However, the results from all pipelines are occasionally discussed, but not reported in the paper. In an effort to be thorough and not record results with unaccounted for weightings, we reached out to the authors to request individual results. Additionally, in other cases, multiple model results from the same pipeline or evolution code are compared and discussed without the authors denoting a preferred result. Rather than arbitrarily selecting a set of results or including all results and biasing the catalog, we reached out to the authors to determine their preferred set of results for inclusion. For all cases where author input was requested, we make note of this in the catalog.

There are two main methods for stellar property inference with asteroseismology. The most convenient method assumes all stars to be scaled versions of the Sun and combines asteroseismic scaling relations with the observed global asteroseismic parameters to provide direct estimates. A more detailed method involves comparing the observed global asteroseismic parameters and/or individual oscillation frequencies with paired stellar evolution and oscillation models. Removing the assumption of solar scaling and, in the case of the individual frequencies, increasing the information content leads to higher achievable precision ($\sim 10-15$\% in age) and accuracy in the inferred properties \citep{chaplin2013}. Although the inclusion of a large number of literature sources where scaling relations factored heavily into the final inferred properties could bias the ages in the catalog, we find that scaling relations are rarely implemented for the Tier 1 and 2 targets. Scaling relations are most often used for fainter stars with low SNR data and/or large samples where a more detailed analysis is less feasible \citep[e.g.,][]{huber2012,lundkvist2014}. We therefore retain the couple of sources where scaling relations are used and note these in the catalog.

We made every effort to include all asteroseismic ages for the Tier 1 and 2 stars present in the literature. However, for $\gamma$ Pav (HD 203608), we intentionally excluded the results from \cite{mosser2008}. \cite{huber2022} determined that the identification of even and odd degree modes was reversed in \cite{mosser2008}, leading to an erroneously high age. In all other cases, the exclusion of any literature source was unintentional.

\subsection{Gyrochronology}\label{sec:meth_gyro}

Our literature search for ages derived via gyrochronology follows largely the same process as that for the asteroseismic ages. We queried ADS using the unique SIMBAD identifiers for each Tier 1 and 2 target along with search terms related to gyrochronology, again removing the requirement for matches to the SIMBAD identifiers for papers published after January 2025. We manually searched the query results to identify ages derived via gyrochronology. We required that ages be derived using measured rotation periods from periodogram analysis to be included in the catalog. While rotation period can be estimated through the chromospheric activity index, $R'_{HK}$, using empirical relations \citep[e.g.,][]{mamajek2008} or using vsini, this significantly increases the uncertainty in the age estimate. Relatedly, we also required that the measured rotation period be traceable to the original source to confirm the method and for reproducibility. For each target present, we record the age, uncertainties (when available), rotation period and original reference, and the reference for the age-rotation relation or method used to derive the age. In cases where multiple empirical relations or methods were used to derive the age, we record all individual results rather than the adopted age, when possible. For stars in binary or multiple systems, we only record the age for component that is the subject of the analysis and not all components of the system.

We made an effort to exclude duplicate ages from the catalog, which was only an issue for literature sources that implemented the pioneering gyrochronology relations of \cite{barnes2007} and \cite{mamajek2008}. The same ($B-V$) and rotation period inputs were often used, resulting in identical ages. In these cases, we elected to include the earlier published age. We note that we did not exclude near duplicates, where the same empirical relation was used with a slightly different ($B-V$) and/or rotation period.

We excluded the gyrochronal ages for 51 Peg (HD 217014) from \cite{maldonado2010} and \cite{vican2012}, which used the rotation period of 37 days from \cite{baliunas1996}. \cite{simpson2010} found a probable period of 21.9 days by reanalyzing an extended sample of the same Mount Wilson HK project data. They only detected a rotation period in the 1998 season data and note that earlier periods from 1980 and 1984 were likely spurious.

\section{Catalog Overview}\label{sec:over}

\begin{figure}[htbp]
\includegraphics[width=0.47\textwidth]{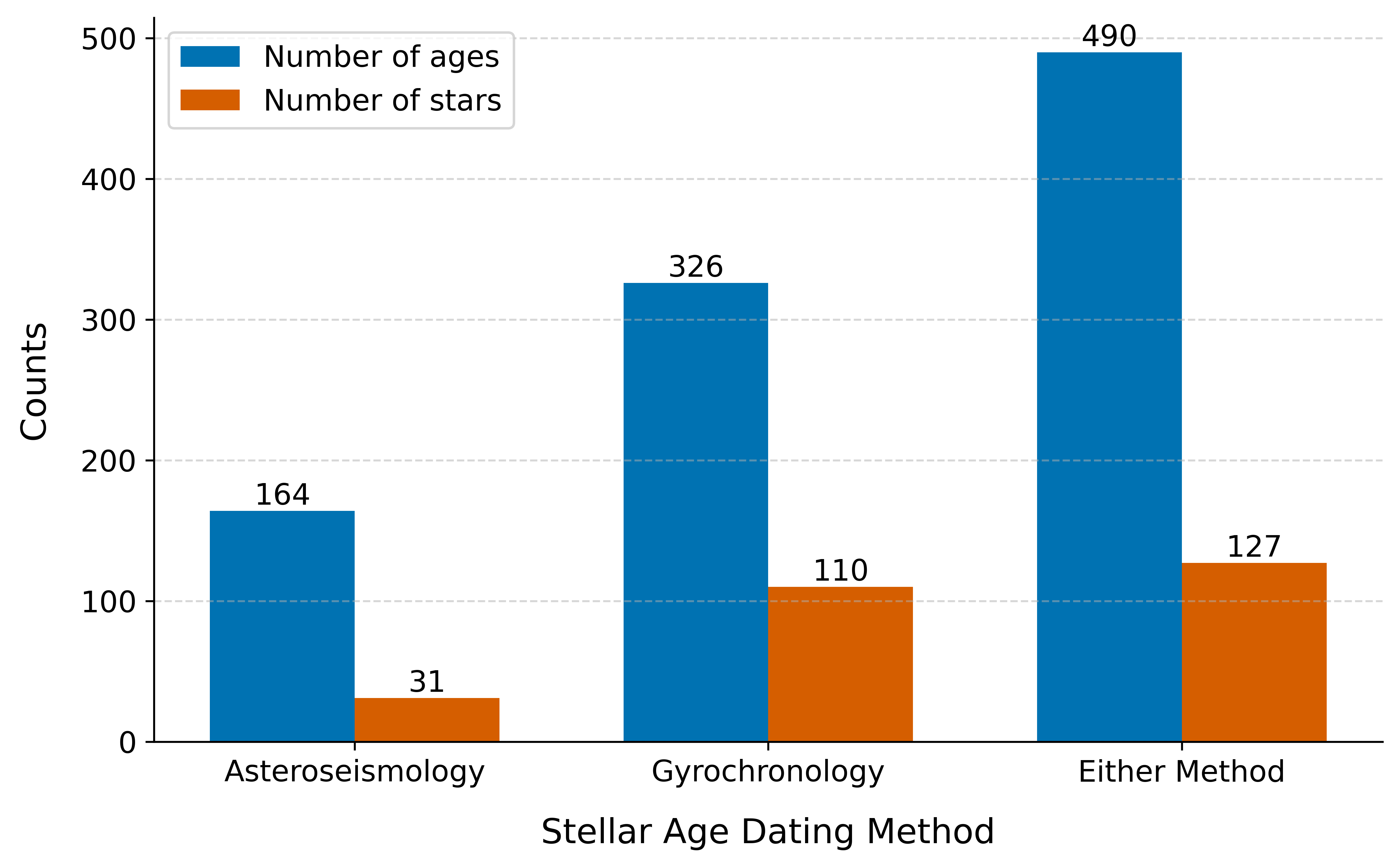}
\caption{Current contents of the HAges Catalog according to stellar age-dating method. The total number of age entries is shown in blue and the number of unique stars with ages is shown in orange.
\label{fig:hages_bar}}
\end{figure}

The HAges Catalog is publicly available at https://doi.org/10.5281/zenodo.19227788 and currently contains stellar ages from 99 literature sources, all of which are listed in Table \ref{tab:lit} of Appendix \ref{sec:app}. From these we extracted a total of 490 stellar ages for 127 stars or $\sim 19$\% of the sample. Figure \ref{fig:hages_bar} shows a breakdown of these ages by the stellar age-dating method. 164 ages were derived through asteroseismology for 31 stars, covering approximately 5\% of the sample with 5 ages per star, and 326 ages were derived through gyrochronology for 110 stars, covering approximately 17\% of the sample with 3 ages per star. Accounting for companions in the catalog to stars with estimated ages would increase these totals. The catalog is separated into 3 tables: one for each stellar age-dating method, which provides the individual published measurements for each star, and an overview table that provides summary statistics for each star and method. Each of the 659 stars in the TSS25 Tier 1 and 2 lists are included in the overview table, regardless of if they have published ages or not, but stars are only included in the method-specific tables if they have a published age. An overview of the content of each of these tables is provided in Table \ref{tab:coldefs} of Appendix \ref{sec:app}.

For each star and method, the overview table includes the average, sample standard deviation, median, spread ($Age_{max}-Age_{min}/2$), and number of individual age measurements. Given the heterogeneous nature of dataset, with age estimates derived using different pipelines, models, and observational data, factoring the reported statistical uncertainties into the average and standard deviation would be non-trivial. The methods for calculating the reported statistical uncertainties themselves vary considerably, ranging from no reported uncertainty or only the model grid interpolation error to detailed Bayesian posterior inference. For these reasons and to show the typical systematic uncertainties, we only calculate the unweighted average and the sample standard deviation of the central values reported. However, we note that the statistical uncertainties of the individual measurements necessarily contribute to this dispersion: the variance of the reported central values contains contributions from both the intrinsic differences between methods and the statistical uncertainties of the individual estimates\footnote{Following from the law of total variance, the variance of the reported estimates can be decomposed into the average statistical variance and the variance of the means \citep[e.g.,][]{casella2024}.}. When only one age from a method is available for a given star, the referenced age is listed as both the average and median, with the standard deviation and spread left blank.

\section{Discussion}\label{sec:disc}

\begin{figure}[htbp]
\includegraphics[width=0.47\textwidth]{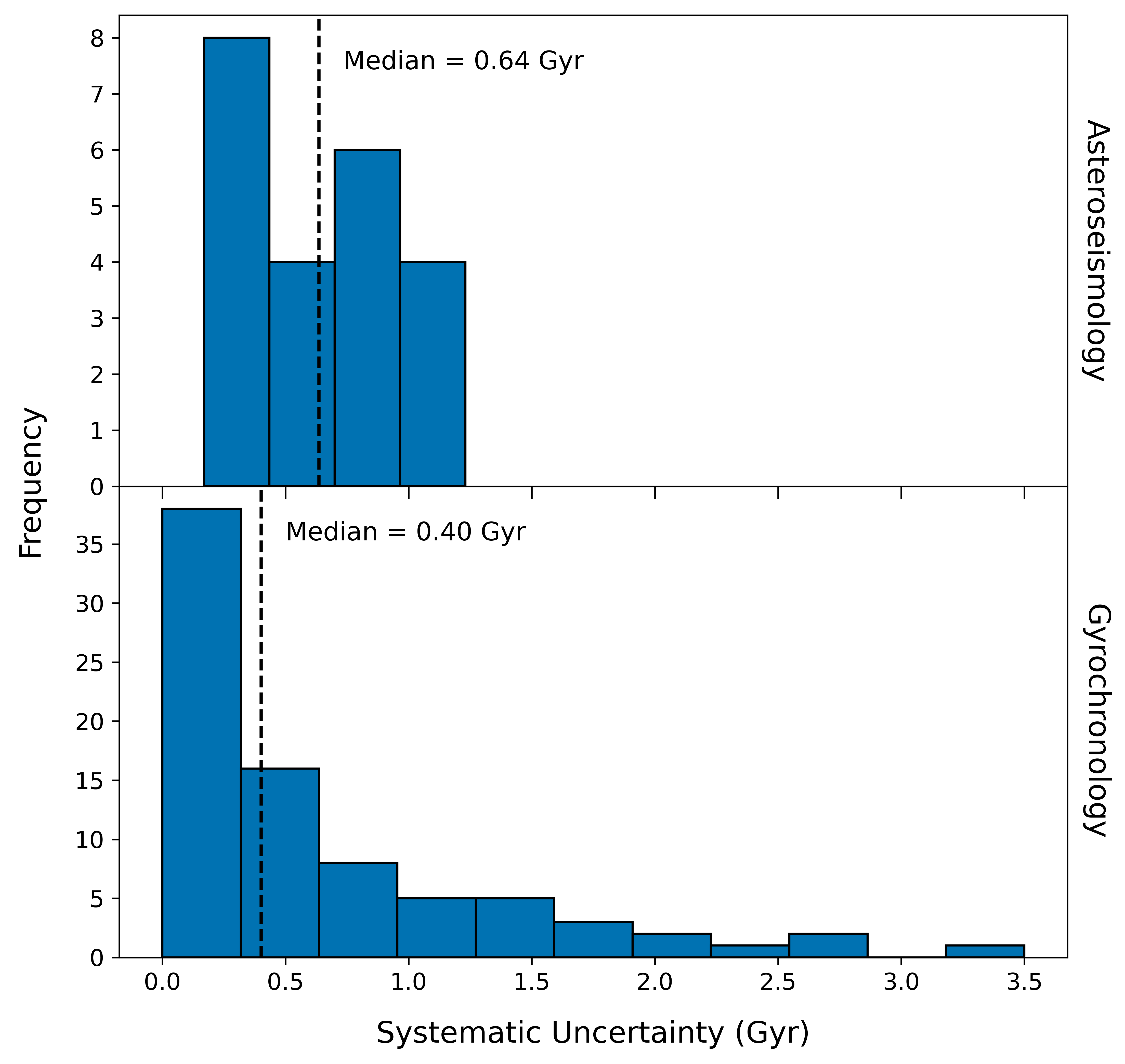}
\caption{Distribution of systematic uncertainties for stars with multiple published ages from asteroseismology (top) and gyrochronology (bottom). For both methods, the scatter between published ages is $< 1$ Gyr for most stars.
\label{fig:hages_hists_gyr}}
\end{figure}

\begin{figure}[htbp]
\includegraphics[width=0.47\textwidth]{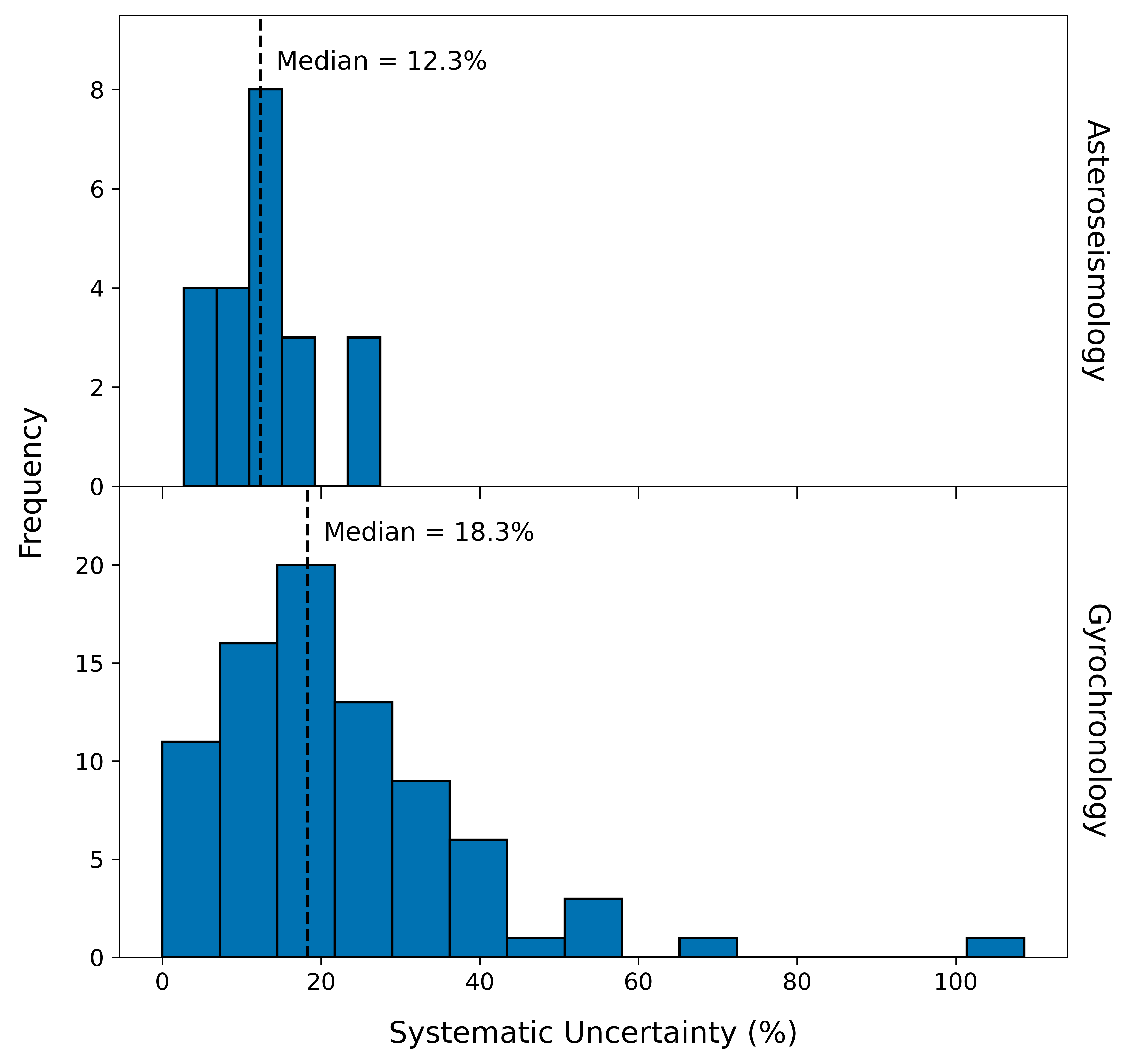}
\caption{The same as Figure \ref{fig:hages_hists_gyr}, but showing the percent systematic uncertainty. Ages from gyrochronology tend to have a higher level of scatter between models, while there is generally excellent agreement between asteroseismic ages with all stars having systematic uncertainties $< 30$\%.
\label{fig:hages_hists_per}}
\end{figure}

A primary goal of the HAges Catalog is to help constrain the systematic age uncertainties for these high priority stars. For stars with multiple published ages from either age-dating method, although a relatively small sample in the case of asteroseismology, the median standard deviation between published ages is $< 1$ Gyr for both methods (Figure \ref{fig:hages_hists_gyr}). The median uncertainty for gyrochronology is significantly less than that for asteroseismology (0.4 Gyr vs. 0.64 Gyr), but this is mainly due to the difference in the sample of ages from these methods, with median ages of 2.09 Gyr and 5.69 Gyr for stars with multiple measurements from gyrochronology and asteroseismology, respectively. The higher precision and level of agreement between the asteroseismic ages are revealed after converting to percent uncertainties, yielding median uncertainties of 12.3\% and 18.3\% (Figure \ref{fig:hages_hists_per}). A significant number of the gyrochronal uncertainties are $> 20$\%, but these are generally for the youngest and oldest stars, where gyrochronology is less precise.

In comparison of the systematic uncertainties in the catalog with the reported uncertainties of individual age measurements, we find that the systematic uncertainties are systematically larger. The medians of the individual reported uncertainties, excluding stars with only one reported age and ages lacking reported uncertainties, are 9\% and 16\% for asteroseismology and gyrochronology, respectively. This highlights the need to catalog published ages and to account for systematic uncertainties when stellar age is an important factor.

As stated in Section \ref{sec:meth_gyro}, we did not exclude near duplicate gyrochronology ages from the catalog, where small differences in the inputs for the same empirical relations result in marginally different ages. This introduces bias into the summary statistics and can lead to erroneously small systematic uncertainties in certain cases. To test the impact of this, we calculated the average and standard deviation of the gyrochronology ages by only including the most recent publications to implement a given empirical relation. We obtain a similar median average age and standard deviation of 2.24 Gyr and 0.37 Gyr (20.2\%). In addition, the median absolute difference between these average ages and those in the HAges Catalog is approximately zero. Therefore, we deem the impact of currently retaining the near duplicate ages to be minimal.

\begin{figure}[htbp]
\includegraphics[width=0.47\textwidth]{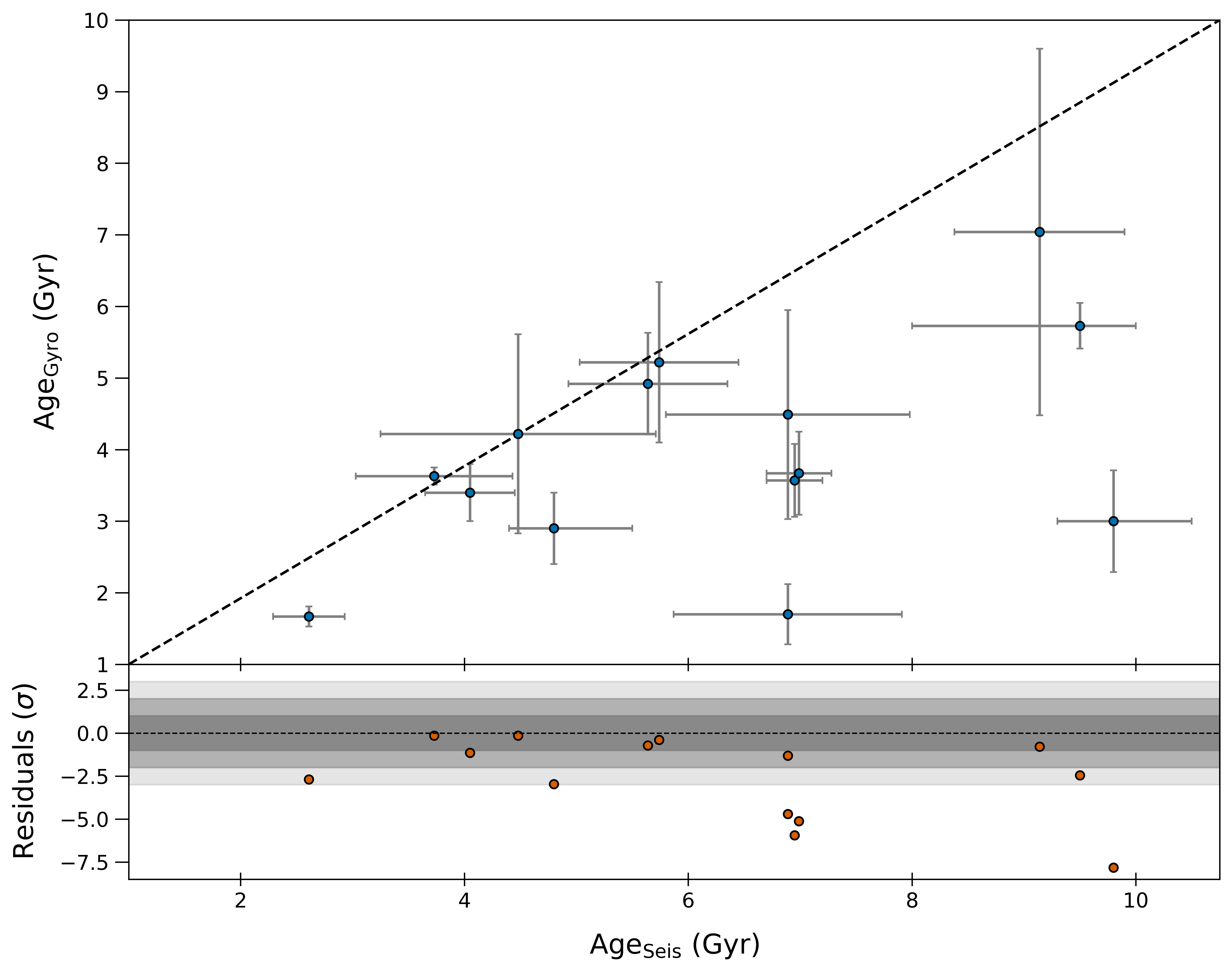}
\caption{Comparison of ages for stars with at least one measurement from both asteroseismology and gyrochronology. For stars with only one measurement from either method, the reported age and statistical uncertainties are shown. Gyrochronology generally agrees with asteroseismology for younger stars, but underestimates stellar ages for stars significantly older than the Sun ($\sim 4.6$ Gyr).
\label{fig:gyro_v_seis}}
\end{figure}

Only 14 stars ($\sim 2$\%) have at least one published age from both asteroseismology and gyrochronology. We show a comparison of the average ages for this sample in Figure \ref{fig:gyro_v_seis}. The asteroseismic and gyrochronal ages are generally consistent for stars $\lesssim 6$ Gyr, with a median difference relative to the asteroseismic age of $\sim -13$\%. For older stars, gyrochronology tends to significantly underestimate the age relative to asteroseismology, with a median difference of $\sim -47$\%. This is expected as older stars undergo weakened magnetic braking and the age-rotation curve flattens.

\subsection{Future Prospects}\label{sec:future}

The lack of precise ages from either age-dating method for $> 80$\% of the sample and that only $\sim 2$\% have ages from both methods highlights the need for a concerted effort to conduct a census of stellar ages over the coming decade(s). A significant increase in the number of asteroseismic ages can likely be expected in the near future as the TESS mission continues to revolutionize the field of time-domain astrophysics \citep{huber2025}. TESS has detected oscillations in hundreds of unevolved stars \citep{hatt2023,lund2025} and $\sim 160,000$ red giants \citep{hon2021}. In cross-matching the HAges Catalog with the recent compilation of bright ($V < 6$) stars with solar-like oscillations detected by TESS from \cite{lund2025}, we find that 133 stars (20\%), including 68 Tier 1 and 65 Tier 2 stars, have available TESS asteroseismic data. Their analysis excluded the known, extremely bright oscillators $\alpha$ Cen A+B and $\alpha$ CMi that require special treatment. Of the 133 stars, 116 do not currently have asteroseismic ages in the catalog. For convenience, we provide a list of the stars with detected oscillations in Table \ref{tab:lund}. We include the total number of asteroseismic ages each star currently has to emphasize those which would benefit most from detailed asteroseismic analysis. With a goal of detecting transiting terrestrial exoplanets orbiting in the habitable zones of bright, Sun-like stars, ESA's PLATO mission, planned to launch in 2026, will spur further work in asteroseismology with long-duration, high-cadence, high-precision photometry for $\sim 1$ million stars \citep{plato}.

\begin{deluxetable}{lclcc}
\tablewidth{0pt}
\tablecaption{TSS25 list stars with detected oscillations from TESS in \cite{lund2025} \label{tab:lund}}
\tablehead{\colhead{SIMBAD Name} & \colhead{TSS Tier} & \colhead{TIC ID} & \colhead{SpT} & \colhead{$N_{ages}$$^a$}}
\startdata
* kap Ret & 1 & TIC 262843771 & F3IV/V          &  \\
*  36 UMa & 1 & TIC 416519065 & F8V             &  \\
* chi Cnc & 1 & TIC 302188141 & F6V             &  \\
* tet Scl & 1 & TIC 70847587  & F5V             &  \\
* rho CrB & 1 & TIC 458494003 & G0+VaFe-1       & 1 \\
*   6 Cet & 1 & TIC 289673491 & F8VFe-0.8CH-0.5 &  \\
*  37 Gem & 1 & TIC 80226651  & G0V             &  \\
HD 114837 & 1 & TIC 255854921 & F6VFe-0.4       &  \\
* alf Men & 1 & TIC 141810080 & G7V             & 1 \\
* nu. Phe & 1 & TIC 229092427 & F9VFe+0.4       &  \\
\enddata
\tablecomments{Table \ref{tab:lund} is published in its entirety in the electronic edition of the {\it Astrophysical Journal}. A portion is shown here for guidance regarding its form and content.\\
\tablenotemark{a} Number of asteroseismic ages in the HAges Catalog.}
\end{deluxetable}

Currently, all of the stars in the HAges Catalog with asteroseismic ages are A-, F-, G-, or K-type stars. Although there is a solid theoretical basis for the presence of oscillations in M dwarfs \citep{palla2005,rodriguez-lopez2012,rodriguez-lopez2014}, no pulsating M dwarfs have been detected to date. A recent analysis of M dwarfs observed by TESS identified a potential candidate, but empirical evidence suggests that detecting oscillations in M dwarfs may require precision approaching 1 ppm that is not reachable with current instrumentation \citep{deamorim2026}. It remains to be seen whether missions like PLATO or monitoring with extreme precision radial velocity spectrographs will enable asteroseismology for the lowest mass stars.

Although a significantly higher proportion of Tier 1 and 2 stars have published gyrochronology ages, more likely have published rotation periods in the literature and current and near future missions could provide rotation periods for the vast majority of potential HWO targets. The TESS All-Sky Rotation Survey has determined rotation periods for $\sim 900,000$ stars within 500 pc \citep{boyle2026}. 30 of the Tier 1 and 2 stars are included in this list, 22 of which do not have gyrochronology ages in the HAges Catalog. The low number of Tier 1 and 2 stars with measured rotation periods from TESS is influenced by the relatively short TESS sectors (27.4 days). Rotation periods from TESS are generally reliable for periods $< 10$ days, but the completeness and accuracy drop significantly at $> 12$ days \citep{boyle2025}. The longer baselines and higher precision of the PLATO mission will aid in measuring rotation periods for many of the slower rotators, with the potential to recover periods for $> 70$\% of observed stars with 6 months and $> 90$\% with 4 years of observations \citep{breton2024}.

Previous age-rotation relations were largely constrained to FGK-type dwarfs younger than the Sun due to a lack of older calibration stellar groups, limited rotation period measurements across spectral types, and the onset of weakened magnetic braking in older stars. Improved sampling of M dwarf rotation periods in stellar groups and companions to higher mass FG-type stars or white dwarfs has enabled extending age-rotation relations down to fully-convective M dwarfs \citep{curtis2020,engle2023,lu2024}. Gyro-kinematic age-dating has been proposed as a means to overcome the flattening of the age-rotation curve at older ages \citep{angus2020,lu2021}. By incorporating the knowledge that the vertical velocity dispersion of stars increases with time due to gravitational interactions, gyro-kinematic relations can infer ages for single stars up to $\sim 14$ Gyr \citep{lu2024}.

While we have focused on these two methods here, we may update the catalog to include other methods capable of age-dating field stars to high precision. One such example is the use of stellar elemental abundance ratios, commonly referred to as "chemical clocks", that trace galactic chemical evolution. In particular, [Y/Mg] and [Y/Al] have been proposed as precise age diagnostics that can achieve precisions of $\sim 1$ Gyr for solar twins \citep[e.g.,][]{dasilva2012,nissen2015,tuccimaia2016,spina2016}, defined as stars having $T_{\rm eff}$ within $\pm 100$ K, log$g$ within $\pm 0.1$ dex, and [Fe/H] within $\pm 0.1$ dex of the Sun \citep{ramirez2014}. Since the number of stars that can be classified as solar twins is low and the bright, nearby solar twins in the TSS25 list are often the stars used to calibrate the [Y/Mg]- and [Y/Al]-age relations, we only found a small number of stars that currently have published ages. However, there are specific circumstances where these relations could prove particularly valuable. The binary solar twins $\zeta^1$ Ret and $\zeta^2$ Ret, which are both Tier 1 stars, were noted to have younger chromospheric and isochronal ages ($\sim 2-5$ Gyr), but Galactic velocity components that imply belonging to the old disk \citep{rochapinto2002,nissen2020}. These types of stars may be blue stragglers formed from the coalescence of low-mass, short-period binaries \citep{rochapinto2002}. \cite{nissen2020} calculated their [Y/Mg] ages to be $9.1 \pm 0.5$ Gyr and $9.4 \pm 0.5$ Gyr for $\zeta^1$ Ret and $\zeta^2$ Ret, respectively, which better matches their kinematic age.

This catalog is publicly available\footnote{https://doi.org/10.5281/zenodo.19227788} and is intended to be a living document that will be updated on an approximately yearly basis with new measurements. If the TSS tiers are updated with new targets, these will be added to the catalog, but we will never remove any stars from the catalog if they are relegated to Tier 3. Finally, we welcome and encourage community feedback on the HAges Catalog to improve its form and content.

%% Please use the acknowledgment and contribution environments. This will 
%% be anonomyized when the "anonymous" style option is used. 
\begin{acknowledgments}

We thank the anonymous reviewer for their timely and thoughtful report that improved the content of this manuscript. We thank Daniel Huber, Marc Hon, and Travis Metcalfe for providing their individual asteroseismology pipeline results. We thank Benard Nsamba, Andreas J{\o}rgensen, Patrick Eggenberger, Andrea Miglio, Mutlu Y{\i}ld{\i}z, Joyce Guzik, Kuldeep Verma, Martin Farnir, and Morgan Deal for helpful input regarding the inclusion of their asteroseismic ages. The results reported herein benefited from collaborations and/or information exchange within NASA’s Nexus for Exoplanet System Science (NExSS) research coordination network sponsored by NASA’s Science Mission Directorate (Grant 80NSSC23K1356, PI Steve Desch). This research has made use of the Astrophysics Data System, funded by NASA under Cooperative Agreement 80NSSC21M0056.

\end{acknowledgments}

\begin{contribution}
%%This section gives authors the space to recognize author contributions. The text inside this environment is NOT counted towards the total word quanta. At a minimum, manuscripts are expected to include this text:

AW was responsible for conducting the literature search, compiling the catalog, and writing and submitting the manuscript. KR provided substantial aid in conducting the literature search and contributed to writing the manuscript. PY provided mentoring and edited the manuscript.

%% But authors are expected to provide more specific details, e.g. 
%%
%%SC was responsible for writing and submitting the manuscript.
%%WWM came up with the initial research concept and edited the manuscript.
%%OTS obtained the funding and edited the manuscript.
%%EBF provided the formal analysis and validation. He also edited the manuscript.
%%GEH Supervised the undergraduates, wrote the software and administers the project github and Zenodo repositories.
%%
%% Authors can use the Contributor Role Taxonomy (CRediT) at
%% https://credit.niso.org
%% for ideas on how write a good statement tailored to their needs.

\end{contribution}

%% To help institutions obtain information on the effectiveness of their 
%% telescopes the AAS Journals has created a group of keywords for telescope 
%% facilities.
%
%% Following the acknowledgments section, use the following syntax and the
%% \facility{} or \facilities{} macros to list the keywords of facilities used 
%% in the research for the paper.  Each keyword is check against the master 
%% list during copy editing.  Individual instruments can be provided in 
%% parentheses, after the keyword, but they are not verified.
%\facilities{HST(STIS), Swift(XRT and UVOT), AAVSO, CTIO:1.3m, CTIO:1.5m, CXO}

%% Similar to \facility{}, there is the optional \software command to allow 
%% authors a place to specify which programs were used during the creation of 
%% the manuscript. Authors should list each code and include either a
%% citation or url to the code inside ()s when available.
\software{astropy \citep{astropy2013,astropy2018,astropy2022},  
          NumPy \citep{numpy}, 
          pandas \citep{pandas},
          ads \citep{ads},
          matplotlib \citep{matplotlib}
          }

%% Appendix material should be preceded with a single \appendix command.
%% There should be a \section command for each appendix. Mark appendix
%% subsections with the same markup you use in the main body of the paper.
%%
%% Each Appendix (indicated with \section) will be lettered A, B, C, etc.
%% The equation counter will reset when it encounters the \appendix
%% command and will number appendix equations (A1), (A2), etc. The
%% Figure and Table counter will not reset.

\appendix

\section{Additional Tables}\label{sec:app}

This section contains additional tables referenced in the main text. Table \ref{tab:lit} lists all literature sources for stellar ages currently included in the HAges Catalog, along with the number of ages adopted from each source and the age-dating method(s) used. Table \ref{tab:coldefs} provides column definitions for all tables in the catalog.

\startlongtable
\begin{deluxetable*}{lcc}
\tablewidth{0pt}
\tablecaption{Literature Sources Included in the HAges Catalog \label{tab:lit}}
\tablehead{
\colhead{Literature Reference} & \colhead{Number of Entries} & \colhead{Method}
}
\startdata
\cite{angus2015} & 15 & Gyrochronology \\
\cite{barnes2007} & 102 & Gyrochronology \\
\cite{bazot2020} & 2 & Asteroseismology \\
\cite{bazot2012} & 1 & Asteroseismology \\
\cite{bazot2016} & 1 & Asteroseismology \\
\cite{bellinger2016} & 2 & Asteroseismology \\
\cite{bonavita2022} & 1 & Gyrochronology \\
\cite{brandao2011} & 1 & Asteroseismology \\
\cite{brandenburg2018} & 1 & Gyrochronology \\
\cite{brandenburg2017} & 24 & Gyrochronology \\
\cite{carrier2005} & 1 & Asteroseismology \\
\cite{castro2021} & 1 & Asteroseismology \\
\cite{chontos2021} & 1 & Asteroseismology \\
\cite{cortez-zuleta2023} & 1 & Gyrochronology \\
\cite{cortes-zuleta2025} & 1 & Gyrochronology \\
\cite{creevey2017} & 2 & Asteroseismology \\
\cite{curtis2020} & 12 & Gyrochronology \\
\cite{deal2021} & 1 & Asteroseismology \\
\cite{delorme2011} & 3 & Gyrochronology \\
\cite{desgrange2023} & 12 & Gyrochronology \\
\cite{donascimento2016} & 1 & Gyrochronology \\
\cite{dogan2010} & 1 & Asteroseismology \\
\cite{dupuy2009} & 2 & Gyrochronology \\
\cite{dupuy2014} & 1 & Gyrochronology \\
\cite{egeland2015} & 1 & Gyrochronology \\
\cite{eggenberger2006} & 2 & Asteroseismology \\
\cite{eggenberger2005} & 1 & Asteroseismology \\
\cite{eggenberger2004} & 2 & Asteroseismology \\
\cite{eggenberger2008} & 6 & Asteroseismology \\
\cite{eisenbeiss2013} & 7 & Gyrochronology \\
\cite{engle2023} & 10 & Gyrochronology \\
\cite{escobar2012} & 3 & Asteroseismology \\
\cite{farnir2020} & 2 & Asteroseismology \\
\cite{filomeno2024} & 1 & Gyrochronology \\
\cite{fouque2023} & 10 & Gyrochronology \\
\cite{gaidos2023} & 13 & Gyrochronology \\
\cite{gray2015} & 1 & Gyrochronology \\
\cite{gruberbauer2013} & 2 & Asteroseismology \\
\cite{grundahl2017} & 1 & Asteroseismology \\
\cite{guenther2004} & 1 & Asteroseismology \\
\cite{guenther2014} & 1 & Asteroseismology \\
\cite{guenther2005} & 1 & Asteroseismology \\
\cite{guzik2016} & 3 & Asteroseismology \\
\cite{hon2024} & 4 & Asteroseismology \\
\cite{huber2022} & 27 & Asteroseismology \\
\cite{johnson2016} & 1 & Gyrochronology \\
\cite{jorgensen2019} & 2 & Asteroseismology \\
\cite{joyce2018} & 2 & Asteroseismology \\
\cite{kajatkari2015} & 1 & Gyrochronology \\
\cite{kayhan2019} & 1 & Asteroseismology \\
\cite{kjeldsen2025} & 1 & Asteroseismology \\
\cite{kuzuhara2013} & 3 & Gyrochronology \\
\cite{lebreton2014} & 1, 2 & Asteroseismology, Gyrochronology \\
\cite{li2019} & 1 & Asteroseismology \\
\cite{li2012} & 1 & Asteroseismology \\
\cite{li2025} & 5 & Asteroseismology \\
\cite{loyd2021} & 1 & Gyrochronology \\
\cite{lundkvist2014} & 5 & Asteroseismology \\
\cite{maldonado2010} & 38 & Gyrochronology \\
\cite{mamajek2012} & 1 & Gyrochronology \\
\cite{mamajek2008} & 8 & Gyrochronology \\
\cite{maxted2015} & 1 & Gyrochronology \\
\cite{metcalfe2015} & 2 & Asteroseismology \\
\cite{metcalfe2023a} & 6 & Asteroseismology \\
\cite{metcalfe2012} & 2 & Asteroseismology \\
\cite{metcalfe2021} & 1, 1 & Asteroseismology, Gyrochronology \\
\cite{metcalfe2023b} & 7 & Asteroseismology \\
\cite{metcalfe2024a} & 5 & Asteroseismology \\
\cite{metcalfe2024b} & 1, 1 & Asteroseismology, Gyrochronology \\
\cite{metcalfe2025b} & 12 & Gyrochronology \\
\cite{miglio2005} & 2 & Asteroseismology \\
\cite{mittag2019} & 1 & Gyrochronology \\
\cite{nielsen2020} & 3 & Asteroseismology \\
\cite{nsamba2022} & 2 & Asteroseismology \\
\cite{pozuelos2020} & 1 & Gyrochronology \\
\cite{quirion2010} & 6 & Asteroseismology \\
\cite{richey-yowell2022} & 2 & Gyrochronology \\
\cite{rieutord2024} & 1 & Asteroseismology \\
\cite{salz2015} & 1 & Gyrochronology \\
\cite{silvaaguirre2017} & 12 & Asteroseismology \\
\cite{soriano2010} & 1 & Asteroseismology \\
\cite{stark2023} & 2 & Gyrochronology \\
\cite{suarez2010} & 1 & Asteroseismology \\
\cite{tang2008} & 1 & Asteroseismology \\
\cite{tang2011} & 1 & Asteroseismology \\
\cite{thevenin2002} & 2 & Asteroseismology \\
\cite{thoul2003} & 2 & Asteroseismology \\
\cite{vauclair2008} & 1 & Asteroseismology \\
\cite{verma2016} & 2 & Asteroseismology \\
\cite{verma2022} & 2 & Asteroseismology \\
\cite{vican2012} & 28 & Gyrochronology \\
\cite{yang2010} & 1 & Asteroseismology \\
\cite{yildiz2007} & 2 & Asteroseismology \\
\cite{yildiz2008} & 2 & Asteroseismology \\
\cite{yildiz2019} & 7 & Asteroseismology \\
\cite{zhang2021} & 1 & Gyrochronology \\
\cite{zurlo2018} & 1 & Gyrochronology \\
\enddata
%\tablecomments{}
\end{deluxetable*}

\startlongtable
\begin{deluxetable*}{lclc}
\tablewidth{0pt}
\tablecaption{Column definitions for HAges Catalog\label{tab:coldefs}}
\tablehead{
  \colhead{Column Name} & \colhead{Unit} & \colhead{Description} & \colhead{TSS25$^a$}
}
\startdata
\multicolumn{4}{c}{\textbf{Columns common to all tables}} \\
\tableline
simbad\_name          &     & Name in SIMBAD database                            & Y \\
TSS\_tier             &     & Highest priority TSS tier that a target belongs to & Y \\
ra                    & deg & Right ascension at epoch J2000                     & Y \\
dec                   & deg & Declination at epoch J2000                         & Y \\
tic\_id               &     & TESS Input Catalog ID                              & Y \\
gaia\_dr3\_id         &     & Gaia DR3 ID                                        & Y \\
hip\_name             &     & Hipparcos ID                                       & Y \\
hd\_name              &     & Henry Draper Catalog ID                            & Y \\
tm\_name              &     & 2MASS ID                                           & Y \\
gj\_id                &     & Gliese designation                                 & Y \\
st\_spectype          &     & Stellar spectral type                              & N$^b$ \\
st\_spectype\_reflink &     & Reference stellar spectral type bibcode            & N$^b$ \\
\tableline
\multicolumn{4}{c}{\textbf{Columns in Overview table}} \\
\tableline
avg\_age\_seis    & Gyr & Average of asteroseismic ages                          & N \\
std\_age\_seis    & Gyr & Sample standard deviation of asteroseismic ages        & N \\
med\_age\_seis    & Gyr & Median of asteroseismic ages                           & N \\
spread\_age\_seis & Gyr & Spread ($Age_{max}-Age_{min}/2$) in asteroseismic ages & N \\
n\_age\_seis      &     & Number of asteroseismic ages                           & N \\
avg\_age\_gyro    & Gyr & Average of gyrochronal ages                            & N \\
std\_age\_gyro    & Gyr & Sample standard deviation of gyrochronal ages          & N \\
med\_age\_gyro    & Gyr & Median of gyrochronal ages                             & N \\
spread\_age\_gyro & Gyr & Spread ($Age_{max}-Age_{min}/2$) in gyrochronal ages   & N \\
n\_age\_gyro      &     & Number of gyrochronal ages                             & N \\
\tableline
\multicolumn{4}{c}{\textbf{Columns in Asteroseismology table}} \\
\tableline
age          & Gyr       & Asteroseismic age                       & N \\
E\_age       & Gyr       & Upper uncertainty in asteroseismic age  & N \\
e\_age       & Gyr       & Lower uncertainty in asteroseismic age  & N \\
mass         & $M_\odot$ & Stellar mass                            & N \\
E\_mass      & $M_\odot$ & Upper uncertainty in stellar mass       & N \\
e\_mas       & $M_\odot$ & Lower uncertainty in stellar mass       & N \\
age\_src     &           & Source of asteroseismic age             & N \\
age\_reflink &           & Reference asteroseismic age bibcode     & N \\
evo\_code    &           & Stellar evolution code used by source   & N \\
osc\_code    &           & Stellar oscillation code used by source & N \\
notes        &           & Additional notes on the literature age  & N \\
\tableline
\multicolumn{4}{c}{\textbf{Columns in Gyrochronology table}} \\
\tableline
age            & Gyr       & Gyrochronal age                        & N \\
E\_age         & Gyr       & Upper uncertainty in gyrochronal age   & N \\
e\_age         & Gyr       & Lower uncertainty in gyrochronal age   & N \\
age\_src       &           & Source of gyrochronal age              & N \\
age\_reflink   &           & Reference gyrochronal age bibcode      & N \\
Prot           & days      & Rotation period used                   & N \\
Prot\_src      &           & Source of rotation period used         & N \\
Prot\_reflink  &           & Reference rotation period bibcode      & N \\
model\_src     &           & Source of age-rotation model used      & N \\
model\_reflink &           & Reference age-rotation model bibcode   & N \\
notes          &           & Additional notes on the literature age & N \\
\enddata
\tablecomments{Columns listed as "common" appear in all tables of the HAges Catalog. The other sections list columns that are specific to each table.\\
\tablenotemark{a} Denotes if the column is taken directly from the TSS25 list.\\
\tablenotemark{b} These columns are taken from the HPIC.}
\end{deluxetable*}

\bibliography{main}{}
\bibliographystyle{aasjournalv7}

%% This command is needed to show the entire author+affiliation list when
%% the collaboration and author truncation commands are used.  It has to
%% go at the end of the manuscript.
%\allauthors

%% Include this line if you are using the \added, \replaced, \deleted
%% commands to see a summary list of all changes at the end of the article.
%\listofchanges

\end{document}